\documentclass{article} 
\usepackage{expl3}
\usepackage{color}
\usepackage[a4paper,vmargin=2.0cm,lmargin=1.0cm,rmargin=5.0cm]{geometry}
\usepackage[pdftex,draft=false,bookmarks]{hyperref}
\usepackage{tabularx}
\usepackage[textwidth=3.5cm]{todonotes}
\usepackage{forarray}
\usepackage{fixme}
\usepackage{etoolbox}
\usepackage{gettitlestring}
\usepackage{thmtools}
\usepackage{amsmath,amssymb}
\usepackage{amsthm}
\usepackage{latexsym}
\usepackage{mathabx} 
\usepackage{manfnt}
\usepackage{tcolorbox}
\tcbuselibrary{listings,theorems}
\tcbuselibrary{skins}
\usepackage[raccourcis]{fast-diagram}
\usepackage{amssymb}
\usepackage[latin1]{inputenc}
\usepackage[T1]{fontenc}
\usepackage{soul} 
\usepackage{pgffor}
\usepackage{hyperref}
\usepackage{mathabx}
\usepackage{MnSymbol}
\usepackage{enumitem}

\definecolor{brightgreen}{rgb}{0.45, 0.95, 0.0}
\definecolor{lightcyan}{rgb}{0.3, 1.0, 1.0}

\usetikzlibrary{arrows}
\usetikzlibrary{positioning,automata,backgrounds}
\usetikzlibrary{calc} 
\usetikzlibrary{trees}
\tikzset{
  treenode/.style = {align=center, inner sep=0pt, text centered,
    font=\rmfamily},
  arn_g/.style = {treenode, circle, draw, 
    text width=1.5em, thick},
  arn_b/.style = {treenode, circle, draw, 
    text width=1.5em, thick},
  arn_r/.style = {treenode, circle, draw, 
    text width=1.5em, thick},
  arn_x/.style = {treenode, circle, black, draw=black,
    text width=1.5em, thick}
}
\tikzstyle{level 1}=[level distance=3.5cm, sibling distance=3.5cm]
\tikzstyle{level 2}=[level distance=3.5cm, sibling distance=2cm]
\tikzstyle{level 3}=[level distance=3.5cm, sibling distance=2cm]
\tikzstyle{bag} = [arn_x,text width=3.0em]

\usepackage{a4}

\begin{document}

\author{Joachim Draeger and Stefan Hahndel}
\title{Simulation-based Unified Risk Assessment for Safety and Security}

\date{\small Technische Hochschule Ingolstadt, Ingolstadt, Germany}

\maketitle





\def\dotcup{\mathbin{\dot\cup}}
\def\cR{{\mathcal{R}}}
\def\real{{\mathbb{R}}}
\def\nat{{\mathbb{N}}}
\def\succ{\mathop{{\mathrm{succ}}}}
\def\post{{{\mathrm{post}}}}
\def\prob{{{\mathrm{Prob}}}}
\def\ldot{\, . \,}
\newtheorem{definition}{Definition}
\newtheorem{remark}[definition]{Remark}
\newtheorem{assumption}[definition]{Assumption}
\newtheorem{lemma}[definition]{Lemma}
\newtheorem{example}[definition]{Example}

\begin{abstract}
The manifold interactions between safety and security aspects 
makes it plausible to handle safety and security risks in an 
unified way. The paper develops a corresponding approach based on 
the discrete event systems (DEVS) paradigm. The simulation-based
calculation of an individual system evolution path provides 
the contribution of this special path of dynamics to the overall 
risk of running the system. Accidentally and intentionally caused 
failures are distinguished by the way, in which the risk contributions
of the various evolution paths are aggregated to the overall risk.  

The consistency of the proposed risk assessment method with
'traditional' notions of risk shows its plausibility. Its 
non-computability, on the other hand, makes the proposed risk 
assessment better suitable to the IT security domain than other 
concepts of risk developed for both safety and security.
Power grids are discussed as an application example and
demonstrates some of the advantages of the proposed method.
\end{abstract}

\section{Introduction}

\subsection{Safety Risks and Security Risks}

%
%
The notion of risk characterizes the expected amount of losses 
associated with the usage of a system $M$. Risk is thus an
important system property. Oddly enough, risk is defined ambiguously. 
It can be characterized from at least two different perspectives, 
safety~\cite{sw2009} and cyber security~\cite{bosworthkabay2002}. 
According to Axelrod~\cite{axelrod2012}, they are distinguished by
who is typically acting on whom, whereby both safety and security 
usually lay down individual requirements on $M$ 
\cite{al2004,firesmith2003,greenetal2015}:
Safety demands that the system must not harm the world; all
deviations from the intended behavior are caused accidentally. 
In the contrary, security demands that the world must not harm the
system, though intelligent adversaries belonging to the world are
acting in an intentionally malicious way. 

Due to these differences, safety and security risk assessments are 
typically executed independent from each other. This may be justified in 
some cases, but may be inappropriate in others. Let us consider 
some examples, in which safety and security risks are intertwined.
\begin{itemize}
\item Let us assume that a decision has to be made whether 
free computational resources of system performance can be invested
either in system monitoring or system defense.  Risk assessments
carried out independently from the safety resp. security perspective 
may hot help in finding an answer.
\item In a cyber attack on a German steel mill in 2014, hackers used 
social engineering techniques for getting access to the control systems 
of the production plant. They modified the control systems in a way, 
that the safety of the plant was compromised. It was not possible 
anymore to shut down a blast furnace. The resulting damage of the 
plant was significant \cite{leeetal2014}. 
\item The Stuxnet worm~\cite{karmouskos2011,my2012} is an example of a
self-propagating malware compromising specific industrial control systems.
As a result, uranium enrichment facilities in Iran seem to suffer
substantial damage.
\end{itemize}
The rapid spread of embedded systems lead to the statement that
there is no safety without security and no security without safety. 
Without a combined view at safety and security, the situations 
described above can not be appropriately analyzed. Instead, 
trade-offs and overlaps between safety and security suggest the 
development of a unified approach to safety and security risk 
assessments as recommended in e.g. \cite{leveson2012}.
%
This paper introduces such a unified notion of risk.

\subsection{Risk Assessment Strategy}

Systematically extending a model of the considered system 
by various safety and security aspects usually leads to a
complex model
(see e.g. \cite{vistbakka2018towards,troubitsyna2018deriving}).
This complexity challenges traditional risk assessment methods 
executed by hand and being informal only.  
For reasons of simplicity, these methods are 
also usually based on static considerations. Static
methods provide results quite fast, they are well applicable to 
systems of significant size, and in many cases the results are a 
sufficiently good approximation to the real situation. In other cases,
however, neglecting system dynamics will be an oversimplification 
\cite{devooght1998,kka2012}. Indeed, \cite{kka2011} states that 
static risk assessments suffer severe limitations as soon as process 
safety is considered. Especially critical in this respect is a
complex dynamics, since minor local fault-related events may 
lead to an unexpected critical global behavior of the overall system
in this way.

Such implications caused by a complex dynamics may have different
roots. Faults may occur concurrently and consecutively and may 
interact with each other. They sometimes propagate across the
system compromising fault control strategies. Back-reactions of 
the system on failure management actions are possible as well.
Intelligent system components like an AI or a human operator 
enable often an effective risk management by their problem 
solving capabilities, but show sometimes an unforseeable behavior. 
If these components serve as the counterpart of an also 
intelligent adversary following an adaptive long-term strategy, 
the risk assessment has to account for planning, learning, 
imperfect decision and other dynamic processes. Static informal 
risk assessments are of limited help in such cases.

%

Consequently, in this paper a simulation-based risk assessment
approach is developed. 
Up to now, the potential of such a risk concept for handling complex
situations is seemingly not yet discussed in necessary depth 
\cite{kpetal2015}. 

\subsection{Related Work}

Despite of the differences between safety and security, an unified
risk assessment is discussed and judged as possible e.g. in 
\cite{brewer1993,jo1992}. Concepts of risk, which are applicable 
to both safety and security, can be found in \cite{axelrod2012,
pcb2013}. Common risk assessment processes, though not simulation-based,
are developed in \cite{madersonstwann,mkm2009}. A concept integrating 
safety and security risks based on fault trees is given in in 
\cite{fm2009}. 

The usefulness of model-based approaches for risk-related 
considerations is shown in \cite{ar2014,arnold1999altarica}.
These models can then be used to simulate different behaviors and
to quantify risk-related properties \cite{pietre2009disentangling}.
Applications of discrete event simulations to cyber security 
problems are discussed in \cite{Chietal2001,felde2010}.
Simulations as tools for risk assessment purposes have been 
discussed in \cite{jk2010} for the special case of stochastically 
varying demands on a production facility. The authors of 
\cite{blometal2006} focus on the Monte-Carlo simulation of air 
traffic control operations. Examples of a simulation-based 
handling of safety without inclusion of security are 
\cite{angskun2007reliability,golabchi2016simulation}. 
Similar considerations from the security risk point of view 
were made in \cite{branagan2012,chi2001network, dissananyaka2011,
nh2007,seo2002modeling,winkelvos2011property}.
A simulation-based analysis of system models from the perspectives 
of both safety and security can be found in 
\cite{bernardi2012dependability,bukowski2016system}.


\subsection{Structure of the Paper}

Section~2 describes, how a system and its potential faults can be 
represented by a formal model. 
In section~3, we start to develop the notion of a simulation-based 
risk measure. At first, this is done for a single individual
evolution of the system. The aggregation of all these risk
contributions provided by the overall set of individual system
evolutions to an overall risk value is described in section~4.
Section~5 demonstrates the advantages of the simulation-based 
risk measure using power grids as an example. 
The paper closes with an outlook discussing key properties of
the proposed risk measure.

\section{Formalization of Systems}

\subsection{Suitability of the DEVS Paradigm}

A formal risk assessment for the system $S$ requires at first a
suitable model of $S$. Such a model can be provided by
the DEVS formalism \cite{zeigler1984,zkp2000} developed 
by Zeigler in 1984. The DEVS formalism is proposed due
to its maturity, generality and flexibility. Its system
definition is closely related to a general (time-dependent) 
system, which assures closeness to practice. DEVS is a 
multi-paradigm formalism
\cite{gon2000theory,praehofer1991system,zeiglerx,zhang2013constructing},
which has the capability to represent (almost) all kinds of 
systems, which have an input/output behavior describable by 
sequences of events \cite{goldstein2013informal,zeigler1976}. It 
can integrate such different system definitions like differential 
equations and discrete event systems in a common framework. This 
supports the handling of complex systems making use of different
system formalizations and being related to different science 
disciplines with individual approaches for describing systems. 
The DEVS paradigm has the expressive power of a Turing machine 
\cite{hg2005}. In principle, it is thus able to represent various 
risk related aspects like risk management actions, risk transfer, 
fault tolerance etc. This property is also helpful for representing 
cognitive aspects, which may be important for the IT security 
perspective. Hence, DEVS models are more general than e.g. 
Bayesian networks or petri nets.


Though the DEVS formalism can handle many types of discrete systems
\cite{zv1993}, it is not able to handle stochastic aspects in its 
original formulation. Safety and security risk assessments are 
inherently stochastical, however, due to the necessity to express 
the frequencies of faults. This gap was closed by the of the STDEVS 
formalism, which is an extension of the DEVS formalism. More precisely,
a DEVS model is a special case of a STDEVS model \cite{kc2006}.

\subsection{DEVS Models of Systems}

In the following, the definition of a DEVS model is recapitulated. 
following \cite{solcany2008,vangheluwe2000}. Being precisely, we will
talk about atomic DEVS models. coupled DEVS models have been defined 
in the literature as well, which are more general from the structural 
point of view. It can be shown, however, that coupled and atomic DEVS 
models have the same expressive power \cite{zeigler1984,zkp2000}.

\begin{definition}[DEVS Model]
An (atomic) DEVS model is an 8-tupel
$M=(X,Y,Q,$
$q_{\mathrm{start}},\delta_{\mathrm{int}},\sigma,\delta_{\mathrm{ext}},\lambda)$
with
\begin{itemize}
\item $X$ as set of input events
\item $Y$ as set of output events
\item $Q$ as set of states
\item $q_{\mathrm{start}}\in Q$ as initial state
\item $\delta_{\mathrm{int}} \colon Q \longrightarrow Q$ 
as the internal transition function 
\item $\sigma \colon Q \longrightarrow \real_0^+ \cup \{\infty\}$
as the time advance function
\item $\delta_{\mathrm{ext}} \colon \bar Q \times 2^X \longrightarrow Q$
as the external transition function defined on
$\bar Q = \{(q,t)\mid q\in Q, 0 \le t \le \sigma(q)\}$ as the total
set of states
\item $\lambda \colon Q \longrightarrow Y \cup \{\phi\}$
as the output function 
\end{itemize}
\end{definition}

\begin{remark}[DEVS Model] \quad\\[-5mm]
\begin{enumerate}[label=\emph{\alph*}),nosep,leftmargin=*]
\item The time advance function $\sigma$ gives the lifetime of
an internal state $q\in Q$. The internal state $q'\in Q$ entered
after reaching the end of the lifetime $\sigma(q)$ of $q$ is determined
by the internal transition function $\delta_{\mathrm{int}}$
via $q'= \delta_{\mathrm{int}}(q)$.
As time in the real world always advances, $\sigma(q)$ must 
be non-negative. The value $\sigma(q)=0$ indicates an instantaneous 
transition. If the system is to stay in an internal state $q$ 
forever, this is modelled by means of $\sigma(q)=\infty$.
\item The definition of the set $\bar Q$ of total states is based on the 
idea to supplement the internal state $q\in Q$ by the elapsed time
$e\in [0,\sigma(q)]$ since the system has entered the state $q\in Q$.
\item External events influence the system as described 
by the external transition function $\delta_{\mathrm{ext}} \colon
\bar Q \times 2^X \longrightarrow Q$. This function can handle sets of 
events representing simultaneously occurring events. 
\item The output event $\lambda(q)$ is generated when the 
time $e$ elapsed after entering the state $q\in Q$ reaches the lifetime 
$\sigma(q)$ of the state $q$, i.e.
$e=\sigma(q)$. At all other times, the output is equal to the non-event $\phi$.
\end{enumerate}
\end{remark}

Incoming events can trigger transitions between states.
Thus, the dynamics of DEVS models is based on the so-called 
time-advance function $\sigma$ and the state transition functions
$\delta_{\mathrm{int}}$ and $\delta_{\mathrm{ext}}$. This leads
to the following description of the dynamics of a DEVS model 
$M=(X,Y,Q,q_{\mathrm{start}},\delta_{\mathrm{int}},\sigma,
\delta_{\mathrm{ext}},\lambda)$ \cite{zeigler2003}.
Let $q\in Q$ be the actual state of $M$. We have to distinguish 
two cases. The first case is that no external event occurs,
the second case handles the arrival of events $x\in 2^X$. In the first
case, the system dynamics is determined by the lifetime $\sigma(q)$ of $q$
and the internal transition function $\delta_{\hbox{int}}$, in the
second case by the external transition function $\delta_{\hbox{ext}}$.

In the first case --- i.e. without the occurrence of external
events $x\in 2^X$ --- the system remains in the state $q$ for time
$\sigma(q)\in\real_0^+ \cup \{\infty\}$. This means:
\begin{itemize}
\item For $\sigma(q)=0$, the state $q$ is immediately changed to the
state $q'\in Q$ given by $q'=\delta_{\hbox{int}}(q)$. This state
transition can not be influenced by external events.
\item For $\sigma(q)=\infty$, the system stays in state $q$ as long as no
external events $x$ occurs.
\item  For $\sigma(q)\in\real^+$, the system outputs the value $\lambda (q)$ 
after expiration of the lifetime $\sigma(q)$ of the state $q$. Afterwards, 
the system 
state changes to $q'\in Q$ given by $q'=\delta_{\hbox{int}}(q)$.
\end{itemize}
In the second case --- i.e. with occurrence of external
events $x\in 2^X$ --- the system changes to a new state
$q'=\delta_{\hbox{ext}}(q,t,x)$, whereby $(q,t)\in \bar Q$ is
the actual total state of $M$ when the set $x$ of events occurs. 


\subsection{STDEVS Models of Systems}

Stochastics is required for representing probabilistically occurring 
safety and security faults. We introduce stochatics by transiting
from the deterministic DEVS formalism to the corresponding probabilistic 
STDEVS formalism. In effect, an (atomic) STDEVS-model is an (atomic) 
DEVS model supplemented by mappings $P_{\mathrm{int}}$, $P_{\mathrm{ext}}$ 
providing transition probability information for the internal and 
external transition functions $\delta_{\mathrm{int}}$,
$\delta_{\mathrm{ext}}$. Thus, an (atomic) STDEVS model has the structure 
\cite{ckw2008,ckw2010}
$M=(X,Y,Q,q_{\mathrm{start}},\delta_{\mathrm{int}},P_{\mathrm{int}},\sigma,
\delta_{\mathrm{ext}},P_{\mathrm{ext}},\lambda)$.
In this definition,
$\delta_{\mathrm{int}} \colon Q \rightarrow 2^Q$ is the internal 
transition function, which describes the set of possible successor
states $\delta_{\mathrm{int}}(q) \subseteq 2^Q$ to the actual state $q$
for situations without occurrence of an external event.
Thus, $\delta_{\mathrm{int}}(q)$
contains all the subsets of $Q$ that the next state can belong to.
The partial function $P_{\mathrm{int}} \colon Q \times  2^Q 
\rightarrow [0,1]$ gives the probability $P_{\mathrm{int}}(q,Q')$ that 
the system model $M$ being in state $q$ makes a transition to a state $q'\in
Q'\in \delta_{\mathrm{int}}(q)$. Concerning the requirements 
for the well-definedness of the probability spaces, 
see \cite{ckw2008,ckw2010}.

Corresponding to  $\delta_{\mathrm{int}}$,
$\delta_{\mathrm{ext}} \colon Q \times \real_0^+ \times 2^X \rightarrow 2^Q$
is the external transition function. It describes the set of possible
successor states $q'\in \delta_{\mathrm{ext}}(q,t,x) \subseteq 2^Q$
for a situation with occurrence of external events $x\in 2^X$, when 
the system model 
$M$ is in a total state $(q,t)\in \bar Q$. Analogous to $P_{\mathrm{int}}$,
the partial function $P_{\mathrm{ext}} \colon Q \times \real_0^+
\times 2^X \times 2^Q \rightarrow [0,1]$ gives the probability 
$P_{\mathrm{ext}}(q,t,x,Q')$ that the system model $M$ being in the 
total state $(q,t)$ makes a transition to a state $q'\in
Q'\in \delta_{\mathrm{ext}}(q)$ at occurrence of events $x$.


For a STDEVS model, the lifetime of a state $q\in Q$ is defined in the same 
way as in the case of a DEVS model, though concerning e.g. safety problems, 
a stochastic lifetime function $\sigma$ would allow a more canonical 
representation of stochastically occurring faults. Being more precise,
the lifetime of a state $q\in Q$ would then become a mapping $\sigma $
from a state to a random variable. If the random variable allows any
time span between two consecutive faults, then the tree of simulation 
paths would contain branching points with uncountably many options 
for a continuation.

\begin{definition}[Language of a STDEVS system]\quad\\
Let $M=(X,Y,Q,q_{\mathrm{start}},\delta_{\mathrm{int}},P_{\mathrm{int}},\sigma,
\delta_{\mathrm{ext}},P_{\mathrm{ext}},\lambda)$
be a STDEVS model and $h\in\real^+_0$ be a nonnegative
real number. The set of possible simulation paths of $M$
limited to the time interval $]0,h]$ is called the {\em language} 
$L(q,h)$ of $M$ for the {\em (time) horizon} $h$ and for the
initial state $q\in Q$. Formally, a simulation path $\tau$ 
is a sequence $\tau=(\rho_1,\ldots,\rho_k)$ representing
the history of the corresponding simulation run consisting of
elements $\rho_j=(q_j,t_j,X_j) \in Q \times \real_0^+ \times 2^X$.
These elements $\rho_j$ document the start resp. end
states of all state transitions $q_{j-1} \rightarrow q_{j}$ during
the simulation run, eventually triggered by the set $X_j$ of 
incoming events. 
In this definition, the start state $q_0$ of the first state 
transition (i.e. $j=1$) is equal to the given initial state $q$.
In the case $q= q_{\mathrm{start}}$, we will usually write 
$L(h)$ instead of $L(q,h)$. 
The language $L(h)$ represents the possible behaviors of the system, 
which can be produced by different faults and event sequences. 
The case $X_j=\emptyset$ indicates 
an internal state transition $q_{j-1} \rightarrow q_{j}$, otherwise 
an external state transition is represented. The times $t_j$ indicate,
how long $M$ was in the state $q_{j-1}$ for $j<k$. For $j=k$, 
the time $t_k$ is limited by the horizon $h$. In this way,
$t_1+\cdots + t_k=h$ is assured. A subsequence
$(\rho_{j},\rho_{j+1},\ldots,\rho_{j'})$ of $\tau$ with 
$1\le j<j'\le k$ is called a {\em subpath} of $\tau$.
\end{definition}

For a DEVS resp. STDEVS model, an event may arrive anytime and may
lead to various state transitions. Though the number of internal 
states in a DEVS resp. STDEVS model is finite and thus countable, of 
course, the set of total states described as a combination of internal 
states and timing information is not. It can be shown, however, that
in a DEVS model these principally uncountable many cases of model 
behavior will only lead to countably many different state transition 
sequences~\cite{hz2006,hz2007}. Since a STDEVS 
model is in essence a DEVS model extended by probabilities of state
transitions, the representing state-transition graph remains finite 
(in an appropriate representation) for a STDEVS model as well. As a 
consequence, the tree of possible state sequences 
of $M$ has a countable size and each node in the tree has 
only a finite number of branching options. For a given finite time 
horizon, the tree of simulation paths is thus finite, too, as long as the
state-transition graph does not contain cycles with transition time
equal to 0. We will assume in the following that such cycles do
not exist in the model $M$.

\begin{remark}[Number of Branching Options]
In the following, we assume that the simulation tree contains
only branching points with a finite number of options. This
condition is fulfilled, if e.g. external events can arrive
only at a finite number of occasions within the time interval 
$[0,h]$. For cases with non-finite many branching options, the
theory, which is presented in this paper, has to be extended. 
This can be done based on the fact that the number of different 
state transitions will remain countable under these circumstances
as well. As soon as the criticalities
assigned to the nodes of the simulation tree depend only on the 
system states and not on timing resp. duration aspects, it will
thus suffice to consider a countable (finite in the case of a
finite horizon) number of sample timings of external events.
If the criticalities depend on timings resp. durations as well,
one may eventually consider the varying arrival times of external 
events via Monte Carlo simulations. 
\end{remark}

Since STDEVS models are a generalization of DEVS models and since the 
expressive power of the DEVS formalism corresponds to that of a Turing 
machine, the class of systems representable by a STDEVS model 
includes all Turing computable situations. Additionally, STDEVS
models cover many types of stochastic discrete systems.

\subsection{Inclusion of Faults in STDEVS Models}

The proposed approach of risk assessment is based on a STDEVS model
$M$ of the system $S$ under consideration. Usually, the model $M$ 
represents only the nominal behavior of $S$. A risk assessment will 
consider off-nominal modes of the system as well, which thus have to
be represented in the model. As a consequence, we need an
extension of $M$ covering safety- and security-related faults
and failures.

In the first step, $M$ is supplemented by components of the 
system environment $U$, which are either affecting the system $S$
or affected by $S$ in a safety or security relevant way. Dependent
on the situations considered as relevant, this may include 
components, which are related to safety and security only in 
an indirect way. Concerning security risk assessments, for example,
the criticality of a violation of the system security will sometimes
depend on the exploitation of this violation. If sensitive data have 
been exposed, the 
attacker may choose the option just to indicate that he has seen these 
data; but he may also use the option to publish these data. The 
criticality of the two choices may be very different.

In the second step, the safety and security problems themselves are 
represented in the model as well as components related to 
problem management.
Especially the adversarial scenario given by cyber security can 
only be handled adequately if both sides --- the attacked 
system $S$ and the attacker --- are modeled at a similar level of 
detail. For example, a cognitive attacker requires a cognitive systems 
control as counterpart for assuring an appropriate defense. Such a 
counterpart keeps track on the attack to avoid unnecessary threats, 
and to organize the defense in an adequate manner. These actions of the
defender are contributing to the controllability of a specific risk 
leading to a mitigation of its criticality.

In the third step, descriptions of the interactions between the 
system $S$ and its environment $U$ are added using the new components,
which are introduced in the first and second step. These interactions
are essential for safety and security considerations as discussed 
in the introduction. 

After these extensions, the model $M$ describes both the nominal and 
off-nominal behavior of the system $S$. Moreover, $M$ is now 
necessarily a stochastic model, since e.g. a fault typically occurs 
with a certain probability. This makes $M$ suitable for
the intended risk assessment. The STDEVS formalism seems to be a 
suitable modeling paradigm for the extended model $M$.

\section{Risk Contributions of Simulation Paths}

\subsection{Simulation Paths as Elementary Risk Contributions}

In the last section, the modeling formalism is described. A simulation
of the resulting model $M$ gives the corresponding system evolution 
with all occurring faults, resulting failures, and their consequences. 
In the following we discuss, how the generated simulation 
history gives the associated contribution to the overall risk.
As usual, the contribution is determined by the criticality assigned
to this specific simulation path --- measuring the amount of 
disadvantages associated with its realization --- and the probability
of its occurrence among all possible system evolutions.
An aggregation of all such risk contributions gives the value of 
the overall risk. From the mathematical point of view, this calculation
defines a risk measure $R$ for an unified assessment of safety and 
security risks. For its formal definition, the technical notions  
of path criticality, path probability, and the path aggregation operator
has to be provided. Before proceeding accordingly, we take a closer 
look at the course of action after the occurrence of a fault. This will
give a better understanding of the dynamical mechanisms associated
with a fault.

We start our considerations with a nominally behaving system. If a component 
of the system starts to behave off-nominal, then the system will usually 
alter the path of dynamics. In a STDEVS model $M=(X,Y,Q,
q_{\mathrm{start}},\delta_{\mathrm{int}}, P_{\mathrm{int}},\sigma, 
\delta_{\mathrm{ext}},P_{\mathrm{ext}}, \lambda)$,
this is represented as a state transition $q_1 \rightarrow q'_1$, 
$q_1,q_1'\in Q$. The new state $q_1' \in Q$ may be the first element
of a state transition sequence, which transmits 
the information about the occurrence of the problem cause --- in the 
following called cause for short ---
to other parts of the system (or its environment). There, the 
consequences of the cause may become effective by executing
another change in the system state, i.e. a state 
transition $q_2 \rightarrow q'_2$, $q_2,q_2'\in Q_2$. Then the new 
state $q_2'$ is the (potentially disadvantageous) effect
of the cause $q_1 \rightarrow q_1'$.
Interpreting a cause as start point of a certain behavior 
the effect can be considered as a (disadvantageous) consequence
of the behavior resulting from the cause.
Such a cause-and-effect resp. causality related perspective of risk is 
discussed in \cite{fmbvt2007,fn}, whereby effects are also called 
consequences. This kind of perspective is supported in \cite{fn2012} 
for safety and in \cite{pirzadeh2011} for security. Additionally,
one has to note that in the description of the general cause-effect 
relationship given above, the state transitions $q_1 \rightarrow q_1'$ 
and $q_2 \rightarrow q'_2$ need not necessarily be different.

\subsection{Criticality of a Simulation Path}

%
%
The representation of system faults, which may contribute to the 
overall risk $R$, in the model $M$ is an important step towards 
actually calclulating $R$, because we are now able to derive the 
{\em existence} of potential problems from $M$. For actually 
evaluating the contribution of this specific problem to the overall 
risk {\em quantitatively}, attributes have to be provided for 
describing its properties. As typical for quantifying a risk, 
one has to know how frequent and how severe a specific system
problem is. The severity is given as criticality $c\colon \bar Q
\rightarrow \real_0^+$ defined on the total states $\bar Q$ of the 
STDEVS model $M$. It measures the amount of disadvantages resulting 
from the occurrence of a specific state $q \in Q$ for a certain 
duration $t\in \real_0^+$. According to this purpose, $c(q,t)\in 
\real_0^+$ will be a nonnegative real number.
States $q$ with $c(q,t)>0$ are representing
modes of the system, which may contribute to the overall risk.

\begin{definition}[Criticality of an Effect]
Let $M=(X,Y,Q,q_{\mathrm{start}},\delta_{\mathrm{int}},P_{\mathrm{int}},\sigma,
		\delta_{\mathrm{ext}},$ $P_{\mathrm{ext}},\lambda)$
be a STDEVS model.
Let $\tau \in L(h)$ be a simulation path of $M$ for the (time) horizon
$h$. The path $\tau=(\rho_1,\ldots,\rho_k)$, $k\ge 1$, with
$\rho_j = (q_j,t_j,X_j)\in Q \times \real_0^+ \times 2^X$ gives the states
$q_j$ together with their lifetimes $t_j$ and thus the total states
$\bar q_j =(q_j,t_j)$. Then the criticality of a total state 
$\bar q_j=(q_j,t_j)$ is given by $c(q_j,t_j)$. Formally, $c$ is a mapping 
$c\colon Q \times \real_0^+ \rightarrow \real_0^+$. In the realm
of criticality, both $q_j$ and $\bar q_j =(q_j,t_j)$
are called an {\em effect}.
\end{definition}

A simulation path $\tau$ may contain many effects
$\bar q_1,\ldots,\bar q_k$. Since these effects $\bar q_j$
can interact with each other, the overall criticality $c(\tau)$ 
of the simulation path $\tau$ may be determined in a more 
complex way than simple summation of the individual 
criticalities $c(\bar q_j)$.
An example would be the disposal of two irritant chemicals. They may
produce a deadly poison in combination \cite{draeger2015}. In other cases,
they may neutralize each other. 
The capability to calculate the overall consequences of several 
failures maybe interacting with and influencing each other is
an important advantage of a simulation-based risk assessment
approach.
As a conclusion, the criticality measure $c$ for simulation paths
must have the potential to take the variety of relationships between 
fault effects into account.
The precise shape of $c$ will thus depend on the specific application.

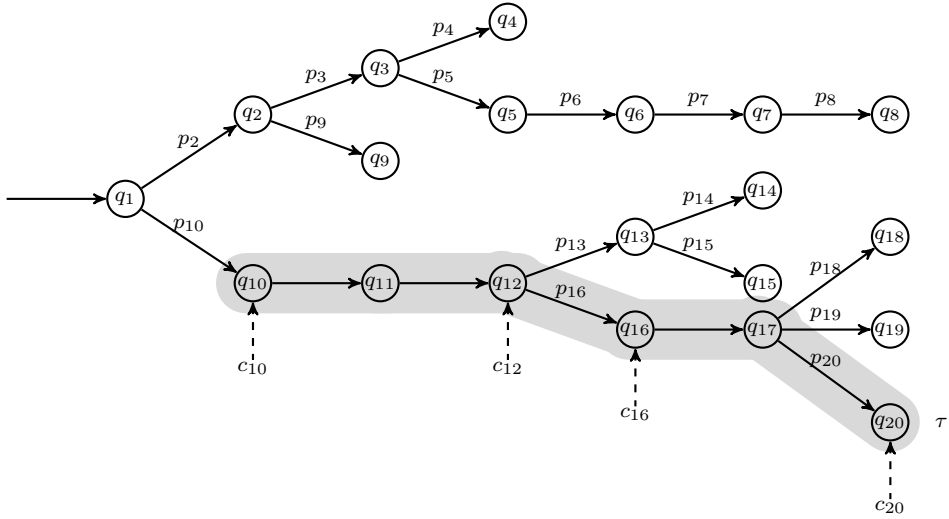
\begin{figure}[h] 
\footnotesize  
\centering  
\begin{tikzpicture}[scale=0.86]
\begin{scope}[grow=east,scale=1.3,yscale=-1,->,>=stealth',level distance=1.5cm, level 1/.style={sibling distance=2cm}, level 2/.style={sibling distance=2cm}, level 3/.style={sibling distance=1.1cm},thick]
\node {} 
child{ node [arn_g] {$q_{1}$} 
  child{ node [arn_r] {$q_{2}$} 
    child{ node [arn_b] {$q_{3}$} 
      child{ node [arn_x] {$q_{4}$} edge from parent node[above] {$p_{4}$}}
      child{ node [arn_x] {$q_{5}$} 
        child{ node [arn_x] {$q_{6}$}
          child{ node [arn_b] {$q_{7}$}
            child{ node [arn_x] {$q_{8}$}
            edge from parent node[above] {$p_{8}$}} 
          edge from parent node[above] {$p_{7}$}}
        edge from parent node[above] {$p_{6}$}}
      edge from parent node[above] {$p_{5}$}} 
    edge from parent node[above] {$p_{3}$}}
    child{ node [arn_x] {$q_{9}$} edge from parent node[above] {$p_{9}$}} 
  edge from parent node[above] {$p_{2}$}}
  child{ node [arn_x] (n1) {$q_{10}$}
    child{ node [arn_x] (n2) {$q_{11}$}
      child{ node [arn_r] (n3) {$q_{12}$}
        child{ node [arn_g] {$q_{13}$}
          child{ node [arn_x] {$q_{14}$} edge from parent node[above] {$p_{14}$}} 
          child{ node [arn_x] {$q_{15}$} edge from parent node[above] {$p_{15}$}} 
        edge from parent node[above] {$p_{13}$}}
        child{ node [arn_x] (n4) {$q_{16}$}
          child{ node [arn_b] (n5) {$q_{17}$}
            child{ node [arn_x] {$q_{18}$} edge from parent node[above] {$p_{18}$}} 
            child{ node [arn_x] {$q_{19}$} edge from parent node[above] {$p_{19}$}} 
            child{ node [arn_x] (n6) {$q_{20}$} edge from parent node[above] {$p_{20}$}} 
          edge from parent node[above] {}}
        edge from parent node[above] {$p_{16}$}}
      edge from parent node[above] {}}
    edge from parent node[above] {}}
  edge from parent node[above] {$p_{10}$}}
edge from parent node[above] {}}; 
\draw[draw = none] (n6) -- ++(0.6,0cm) node {$\tau$};
\draw[<-,black, shorten >=1mm, dashed] (n1) -- ++(0,1cm) node {$c_{10}$};
\draw[<-,black, shorten >=1mm, dashed] (n3) -- ++(0,1cm) node {$c_{12}$};
\draw[<-,black, shorten >=1mm, dashed] (n4) -- ++(0,1cm) node {$c_{16}$};
\draw[<-,black, shorten >=1mm, dashed] (n6) -- ++(0,1cm) node {$c_{20}$};
\end{scope}
\begin{scope}[on background layer]
  \draw[line width=0.8cm,cap=round,join=round,black!15, fill=black!8]
  ($(n1)!-0.1cm!(n2)$) -- ($(n1)!2.1cm!(n2)$);
    \draw[line width=0.8cm,cap=round,join=round,black!15, fill=black!8]
  ($(n2)!-0.1cm!(n3)$) -- ($(n2)!2.1cm!(n3)$);
    \draw[line width=0.8cm,cap=round,join=round,black!15, fill=black!8]
  ($(n3)!-0.1cm!(n4)$) -- ($(n3)!2.1cm!(n4)$);
    \draw[line width=0.8cm,cap=round,join=round,black!15, fill=black!8]
  ($(n4)!-0.1cm!(n5)$) -- ($(n4)!2.1cm!(n5)$);
    \draw[line width=0.8cm,cap=round,join=round,black!15, fill=black!8]
  ($(n5)!-0.1cm!(n6)$) -- ($(n5)!2.4cm!(n6)$);
\end{scope}
\end{tikzpicture}
\caption{\label{treefigure}
The figure shows the progressively diversifying state tree produced by 
the simulation of a stochastic model. The simulation path $\tau$ 
contains several disadvantageous consequences occurring in the 
states $q_{10}$, $q_{12}$, $q_{16}$, and $q_{20}$. These disadvantages
are quantified by the criticalities  $c_{10}$, $c_{12}$,
$c_{16}$, and $c_{20}$. When assessing the overall criticality $c(\tau)$,
all the $c_{10}$, $c_{12}$, $c_{16}$, and $c_{20}$ have to be taken into
account and calculated with each other.
}
\end{figure}

\begin{definition}[Criticality of Effects]
Let $M=(X,Y,Q,q_{\mathrm{start}},\delta_{\mathrm{int}},P_{\mathrm{int}},\sigma,
		\delta_{\mathrm{ext}},P_{\mathrm{ext}},$ $\lambda)$
be a STDEVS model.  Let $\tau \in L(h)$ be a simulation path of 
$M$ for the (time) horizon $h$. 
The path $\tau=(\rho_1,\ldots,\rho_k)$, $k\ge 1$, with
$\rho_j = (q_j,t_j,X_j)\in Q \times \real_0^+ \times 2^X$ gives the states
$q_j$ together with their lifetimes $t_j$ and thus the total states
$\bar q_j = (q_j,t_j) \in \bar Q$. For handling multiple faults,
the domain of $c$ consists of a (temporally ordered) sequence 
$\bar q=(\bar q_1,\ldots, \bar q_k)$ of individual total states.
Thus, the extended criticality $c$ has the signature 
$c\colon \bar Q\times \cdots \times \bar Q \rightarrow \real_0^+$. 
\end{definition}
The definition above extends the criticality $c$
in such a way, that criticality correlations can be taken into account 
(see figure~\ref{treefigure}).
The lifetimes $t_j$ of the total states $\bar q_j$ provide information
about time differences between the effects, which may influence $c$ as 
well. If the criticality correlation depends on additional parameters,
the values of these parameters can typically be coded in the states $Q$
of a STDEVS model.

\subsection{Probability of a Simulation Path}

\begin{figure}[h] 
\begin{center}
\begin{tikzpicture}[grow=right, sloped, ->, scale=1.3,>=stealth',thick]
\node[bag, label=right: {\hspace*{3.35cm} \Huge $\vdots$}] {$q_{\mathrm{start}}$}       
    child {
        node[bag, label=right: {\hspace*{3.35cm} \Huge $\vdots$}] {$q_{m}$} 
            child {
                node[bag, label=right:
                    {$\tau_{ms}$}, label=above: {}] {$q_{ms}$}
                edge from parent
                node[above] {$p(q_m \rightarrow q_{ms})$}
            }
            child {
                node[bag, label=right:
                    {$\tau_{m1}$}, label=above: {}] {$q_{m1}$}
                edge from parent
                node[above] {$p(q_m \rightarrow q_{m1})$}
            }
        edge from parent 
            node[above] {$p(q_{\mathrm{start}} \rightarrow q_{m})$}
    }
    child {
        node[bag, label=right: {\hspace*{3.35cm} \Huge $\vdots$}] {$q_{1}$}        
        child {
                node[bag, label=right:
                    {$\tau_{1r}$}, label=above: {}] {$q_{1r}$}
                edge from parent
                node[above] {$p(q_1 \rightarrow q_{1r})$}
            }
            child {
                node[bag, label=right:
                    {$\tau_{11}$}, label=above: {}] {$q_{11}$}
                edge from parent
                node[above] {$p(q_1 \rightarrow q_{11})$}
            }
        edge from parent         
            node[above] {$p(q_{\mathrm{start}} \rightarrow q_{1})$}
    };
\end{tikzpicture}
\end{center}
\caption{ \label{bayesfigure}
The simulation of a determninistic model gives a single sequence $\tau$ 
of system states. For stochastic models, the state sequence diversifies
to a tree of possible simulation paths. The probability of transiting
to a specific successor state at a branching point in the tree is determined 
by the probability $P_{\mathrm{int}}(q',\{q''\})$ assigned to the corresponding
state transition $q'\rightarrow q''$. Let us take a closer look at the 
simulation path $\tau_{11}$ representing the state sequence
$q_{\mathrm{start}} \rightarrow q_1 \rightarrow q_{11}$. Using the
abbreviations $T := q_{\mathrm{start}} \rightarrow q_1$ and 
$T':= q_1 \rightarrow q_{11}$, the probability $p(\tau_{11})$ of the 
occurrence of path $\tau_{11}$ is equal to
the probability $p(\tau_{11}) = p(T\wedge T')$ that both state 
transitions $T,T'$ occur. Applying Bayes rule, it holds $p(T\wedge T') =
p(T)\cdot p(T\mid T')$. In the example, $p(T)=p(q_{\mathrm{start}}
\rightarrow q_1)$ is the probability that the state $q_1$ is reached 
from the start state $q_{\mathrm{start}}$. The probability $p(T\mid T')$
on the other hand is the probability that from the state $q_1$, which have
been reached after execution of $T$, a transition to the state $q_{11}$
takes place. This means $p(T\mid T') = p(T')= p(q_1 \rightarrow q_{11})$.
}
\end{figure}
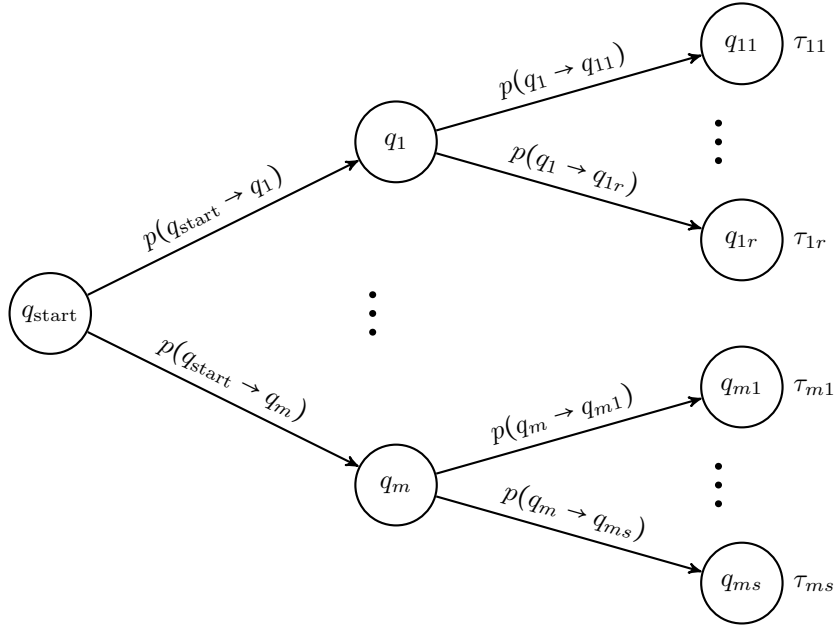

Safety and security problems will occur probabilistically.
Accordingly, the overall dynamical behavior of a system model $M$ 
displays a tree instead of a single path. The probability of taking 
a specific branching option in this tree is given by the probability
$p(\gamma)$ of the corresponding state transition $\gamma$.
For calculating the probability $p(\tau)$ of a
whole simulation path $\tau$, which may result from several such 
branching choices $\gamma_i$, we have to compose the probabilities 
$p(\gamma_i)$ assigned to these choices $\gamma_i$.
This can be done with the Bayes rule (see 
figure~\ref{bayesfigure}). 

\begin{definition}[Probability of Cause]
Let $M=(X,Y,Q,q_{\mathrm{start}},\delta_{\mathrm{int}},P_{\mathrm{int}},\sigma,
		\delta_{\mathrm{ext}},P_{\mathrm{ext}},$ $\lambda)$
be a STDEVS model. Let $\gamma = (q, t, X', q')$ be a state transition 
$q \rightarrow q'$ between states $q,q'\in Q$ occurring at lifetime $t$ 
of state $q$, eventually triggered by a set $X'$ of external events 
($X'=\emptyset$ is a valid choice).
Then the probability of executing $\gamma$
is designated as $p(\gamma)$. The value of $p(\gamma)$ is given by 
the internal transition probability $P_{\mathrm{int}}(q,\{q'\})$ for
$X'=\emptyset$ and by the external transition probability 
$P_{\mathrm{ext}}(q,t,X',\{q'\})$ for $X'\neq\emptyset$ with
the system being in the total state $(q,t)$.
The 4-tupel $\gamma$ represents a so-called {\em cause}.
\end{definition}

\begin{definition}[Probability of a Sequence of Causes] \label{probsequdef}
Let $M=(X,Y,Q,q_{\mathrm{start}},\delta_{\mathrm{int}},$ $P_{\mathrm{int}},
\sigma, \delta_{\mathrm{ext}},P_{\mathrm{ext}},\lambda)$ be a STDEVS model. 
Let $\tau \in L(h)$ be a simulation path of $M$, with (time) horizon $h$. 
Assigned to $\tau=(\rho_1,\ldots,\rho_k)$ with
$\rho_j = (q_j,t_j,X_j)\in Q \times \real_0^+ \times 2^X$ is the
(temporally ordered) sequence $\gamma =(\gamma_1,\ldots,\gamma_{k})$ of 
state transitions $\gamma_j := (q_{j-1},t_j,X_j,q_{j})$ with
$q_0:= q_{\mathrm{start}}$. Then the probability $p(\gamma)$ of the 
occurrence of the sequence $\gamma$ is given by
$$p(\gamma)= p(\gamma_1)\cdot p(\gamma_1\mid \gamma_2)\cdot 
	\cdots \cdot p(\gamma_1,\ldots,\gamma_{k-1} \mid \gamma_k)$$
according to Bayes rule. The expression $p(\gamma_1,\ldots,\gamma_{j-1}
\mid \gamma_j)$ results from the fact that when the state transition 
$\gamma_j$ is triggered, the state transitions 
$\gamma_1,\ldots,\gamma_{j-1}$ were already executed and have set
the preconditions for $\gamma_j$. 
\end{definition}

\subsection{A Notion of Risk for Simulation Paths}

We now define risk contribution provided by an individual system
behavior represented by a corresponding 
simulation path $\tau = (\rho_1,\ldots,\rho_k)\in L(h)$. Using the
causes $\gamma = (\gamma_1,\ldots,\gamma_k)$ and the effects $\bar q = 
(\bar q_1,\ldots,\bar q_k)$ belonging to the path $\tau$, we
are now able to assign both a probability and a criticality to $\tau$ via 
the measures $p(\gamma)$ and $c(\bar q)$ defined in the last section.

\begin{definition}[Probability and Criticality of Simulation Paths]~\\
Let $M$ be a STDEVS model and $h\in\real^+_0$ the horizon of the 
simulation.  Let $\tau \in L(h)$ be a simulation path of $M$ for 
the (time) horizon $h$. 
The path $\tau = (\rho_1,\ldots,\rho_k)$ 
is associated with a sequence $\gamma= (\gamma_1,\ldots,\gamma_{k})$ 
of causes and a sequence $\bar q = (\bar q_1, \ldots,
\bar q_k)$ of effects. Then
the probability $p(\tau)$ and the criticality $c(\tau)$ of the path
$\tau$ are defined as $p(\tau):=p(\gamma)$ and $c(\tau):=c(\bar q)$. 
\end{definition}

Using the probability $p(\tau)$ and criticality $c(\tau)$ of the 
simulation path $\tau$ we will now define a risk measure $R$ for 
a path $\tau$. 

\begin{definition}[Risk Measure for Simulation Paths] \label{rmofpaths}
Let $M$ be a STDEVS model and $h\in\real^+_0$ be a horizon.
Then a risk measure $R\colon L(h)\longrightarrow \real^+_0$ can be 
defined for the simulation paths $\tau\in L(h)$ of $M$ by
assigning a nonnegative real value to $\tau$ defined by $R(\tau)
= p(\tau)\cdot c(\tau)$.
\end{definition}

\section{Overall Risk of a System}

\subsection{Overall Risk as Aggregation of Risk Contributions}

A simulation can construct a tree-like representation of the system
behavior consisting of individual imulation paths. This representation 
contains both safety and security problems. For safety problems, the
probabilities assigned to branching options are usually determined
locally. For security problems, the story may be different. Intelligent
attackers (and defenders as well) may predict the outcomes of the
various simulation paths and then they will select the most 
promising one for finding the best way to act. Then, the probabilities
of choosing specific consecutive branching options are not statistically
independent anymore. It is beyond the scope of this paper to describe,
how these probabilities are determined. 
According to \cite{suzuki2013solution}, stochastics and strategies 
has to be integrated in the context of stochastic 
game theory \cite{mertens1981stochastic,neyman2003stochastic}.  
Instead, we will focus on the concept, how the risk contributions 
$R(\tau)$ provided by the set $L(h)$ of possible simulation paths 
$\tau$ for a given time horizon $h$ are aggregated to a risk assessment 
$R$ for the model $M$. Since $L(h)$ describes the overall behavior of 
$M$, it is plausible to define a risk measure $R$ for $M$ as sum over 
the risk values $R(\tau)$ assigned to the different simulation paths 
$\tau \in L(h)$ of $M$. In this way, the risk $R$ assigned to $M$ is
the sum of the criticalities $c(\tau )$ of the paths $\tau \in L(h)$
weighted by their probabilities $p(\tau )$. This corresponds to the 
traditional form of a risk measure for safety aspects as expectation 
value of the criticality over all possible cases. 

\begin{definition}[Risk Measure] \label{defriskclass}
Let $M=(X,Y,Q,q_{\mathrm{start}},\delta_{\mathrm{int}},
P_{\mathrm{int}},\sigma, \delta_{\mathrm{ext}},P_{\mathrm{ext}},\lambda)$
be a STDEVS model and $h\in\real^+_0$ be a horizon.
Then a risk measure $R\colon \real^+_0 \longrightarrow 
\real^+_0$ parameterized by the horizon $h$ is defined by
$$R(h) :=\sum\limits_{\tau\in L(h)} R(\tau)= 
\sum\limits_{\tau\in L(h)} p(\tau)\cdot c(\tau) $$
If we consider a language $L(q,h)$ for the initial state $q\in Q$ 
instead of $L(h)$ for the canonical choice $q=q_{\mathrm{start}}$,
the corresponding risk is designated as $R(q,h)$.
\end{definition}

The definition of $R(h)$ can be considered as plausible, because
a limitation of the horizon $h$ reduces the definition to 
traditional definitions of e.g. safety risk. This topic is discussed
more thoroughly in the outlook. 

\begin{remark}[Mixed Random/Strategic Situations] 
The necessity of a unified handling of stochastics and strategies may
not be limited to considerations regarding cyber security.
As soon as the system contains a cognitive control component following
a long-term aim, the control actions chosen along a simulation
path are not uncorrelated anymore. This situation is analogous
to the case of a cognitive attacker, who is trying
to exploit a system not actively defended. For stochastic
systems, in which only one side is following a long-term aim, 
the general theory of stochastic games is not required. Instead,
representing the situation as Markov decision process will suffice.
\end{remark}

\subsection{Correlation of Probabilities by Cognitive Entities}

In the preceeding section we have defined
a simulation-based risk measure unifying safety and security.
The risk measure is determined by the criticalities and 
probabilities of the simulation paths. The underlying idea
should be clear from the perspective of safety. For security,
however, it may be not immediately clear how the existence of
cognitive entities will lead to a correlation of probabilities
due to their individual aims, strategies and long-range plans.
For explaining this effect, we have to give several definitions 
at first.

\begin{definition}[Simulation Path Operations] \quad\\[-5mm]
\begin{enumerate}[label=\emph{\alph*}),nosep,leftmargin=*]
\item The operator $\circ$ designates the concatenation of two
	simulation paths
\item Let $\tau =(\rho_1,\ldots,\rho_k) \in L(q_{\mathrm{start}},h)$
with $\rho_j=(q_j,t_j,X_j)$
be a simulation path and $q_l$ with $1<l<k$ a system state 
occurring on $\tau$. Let $\tau_1=(\rho_1,\ldots,\rho_{l-1})$
be the subpath of $\tau$ from $q_{\mathrm{start}}$ to $q_{l-1}$.
Then it exists a simulation path $\tau_2\in L(q_l,h')$ for 
the horizon $h' = h - \sum_{j=1}^{l-1} t_j$ 
with $\tau=\tau_1\circ \tau_2$. In the following,
we will use the notation 
$\tau_{\post}(\rho_l) := \tau_2$.
\end{enumerate}
\end{definition}

\begin{definition}[Subsets of a Language] \quad\\[-5mm]
\begin{enumerate}[label=\emph{\alph*}),nosep,leftmargin=*]
\item Let $L_{\tau_1}(h)\subseteq L(h)$ designate the subset of all
paths $\tau = (q_j,t_j,X_j)_{j=1}^k \in L(h)$, which start with a
common subpath $\tau_1 = (q_j,t_j,X_j)_{j=1}^l$, $l\le k$, 
of $\tau$. This means, that for a path $\tau\in L_{\tau_1}(h)$ it exists 
a path $\tau_2 \in L(q_{l+1},h')$ for the
horizon $h' = h - \sum_{j=1}^l t_j$ with $\tau=\tau_1\circ \tau_2$.
\item Let $\rho$ be a node in the simulation tree. Then $\succ(\rho)$ 
designates the set of nodes, which succeeds the node $\rho$ in a 
path $\tau\in L(h)$. For all members $\rho'\in\succ(\rho)$ with
$\rho=(q,t,X)$, $\rho'=(q',t',X')$ exists an internal or external 
state transition from $q$ to $q'$.
\end{enumerate}
\end{definition}

After providing the necessary notational definitions, we will now 
discuss what happens in a decision point of the simulation tree. 
Let the system be in the state $q\in Q$ in this decision point.
From start state $q_{\mathrm{start}}$ to decision point the 
simulation has already generated the path $\tau_1\in 
L(q_{\mathrm{start}},h)$ for a horizon $h$. In general, the decider
will include the path $\tau_1$ in her considerations, because 
the overall criticality of a simulation path $\tau$ may very well 
depend on events occuring in the subpath $\tau_1$.

A cognitive entity, which is responsible for making the decision 
in state $q$, determines the transition probabilities 
$\prob(q\rightarrow q')$ of the possible continuations given by 
$\rho'\in\succ(\rho)$ with $\rho=(q,t,X)$, $\rho'=(q',t',X')$
according to its decision. The set $\succ(\rho)$ represents the
available choices of the pending decision. A decider acting totally
rational and faultless may using only yes/no-decisions (i.e.
$\prob(q\rightarrow q')\in \{0,1\}$). As soon as imperfections of
the decision process are taken into account, the probabilities may
also assume intermediate values. 
The decision process itself is of no relevance for risk assessment. 
Furthermore, a simulation tree developed by the decider for 
predictive purposes may usually differ from the corresponding part 
of the simulation tree used for the risk calculation. 


\subsection{Special Case of (Risk-)Rationality}

We supplement our considerations with some remarks concerning a
situation, in which both attacker and defender ---
the two deciders belonging to the considered system --- are using 
the same simulation tree as the risk assessment procedure.
They are following the explicit goals of risk maximization 
and risk minimization, respectively.
It results an adversarial situation. The win of one 'player'
is the loss of the 'other'. If we additionally assume as a
simplification that the system is strictly deterministic besides
of the decisions to be made and that attacker and defender are 
executing measures 
and countermeasures alternately, the description as a (combinatorial) 
zero-sum game becomes adequate {\cite{bier2008game}}. 
Then, the definition of the overall risk $R$ can be based
on a minimax algorithm {\cite{rn2013}}, which is processing
the simulation tree recursively. Executing a recursive minimax
algorithm instead of just summing up the risk contributions 
assigned to the indivifual simulation paths is the result of
integrating the decisions of attacker and defender on the one hand
and the risk assessment procedure on the other.

We discuss the situation at a specific node $\rho=(q,t,X)$ of the 
simulation tree. We will define the risk inductively. 
Let $\tau_1$ designate the simulation path from 
the root node of the simulation tree to $\rho$. Let us suppose 
for a moment that no decision has to be made on the pathway $\tau_2$
from $\rho$ to the terminating leaf in the simulation tree. Since
the system is assumed to be deterministic, this condition means that 
$\tau_2$ does not contain a branching point. Thus, 
$L_{\tau_{\post}(\rho)}(h) \subseteq L(h)$ consists of a single path 
$\tau_1\circ\tau_2$ only. The risk assigned to this path 
(see definition~\ref{rmofpaths}) is equal to the risk assigned
to $L_{\tau_{\post}(\rho)}(h) = \{\tau_1\circ\tau_2\}$.

If a decsion has to be made in the node $\rho$, it exists more than
one possible continuation. The corresponding set $L_{\tau_{\post}(\rho)}(h)$ 
of simulation paths has the structure
$$L_{\tau_{\post}(\rho)}(h) = 
\bigcupdot_{\rho'=(q',t',X')\in \succ(\rho)} L_{\tau_{\post}(\rho')}(h-t').$$
Based on the induction hypothesis, the risk $R_{q'}$ assigned to 
$L_{\tau_{\post}(\rho')}(h)$ is already known. For
calculating the risk $R_q$ assigned to $L_{\tau_{\post}(\rho)}(h)$,
the definitions \ref{probsequdef} and \ref{defriskclass} lead to
$$R_q = \sum\limits_{q'\in\succ(\rho)} \prob(q\rightarrow q') R_{q'}.$$
It remains to determine the transition probabilities representing
the decision result. For convenience, let us designate 
$R^{\max}(q,h) := \max\limits_{q'\in\succ(\rho)} R(q',h)$
and
$R^{\min}(q,h) := \min\limits_{q'\in\succ(\rho)} R(q',h)$.
In both cases exist a node $\rho'\in\succ(\rho)$ succeeding $\rho$ 
in a simulation path with $R^{\max}(q,h) = R(q',h)$
resp. $R^{\min}(q,h) = R(q',h)$. This state is
designated as $q'_{\mathrm{max}}$ resp. $q'_{\mathrm{min}}$. 
We assign the transition probability $\prob(q\rightarrow q') =1$ 
for $q'= q'_{\mathrm{max}}$ resp. $q'= q'_{\mathrm{min}}$ and
$\prob(q\rightarrow q') =0$ otherwise. 

The induction stops when the root $q_{\mathrm{start}}$ 
of the simulation tree is reached. For $q_{\mathrm{start}}$, it holds
$R(h) = R_{q_{\mathrm{start}}}$.


\section{Example Power Grids}

\subsection{Power Grids as Exemplary Application}

Though the proposed approach of an unified assessment of safety and 
security risks is appealing from the theoretical point of view,
a systematic processing of all possible evolution paths will require
a significant computational effort. The necessary effort is justified, 
however, if the system under consideration has e.g. a complex dynamics
hardly accessible by static evaluations. Distribution networks like 
power grids \cite{pcp2013} have this property due to phenomena like 
cascading failures. Additionally, they can be modeled canonically in
a very simple way as a network.
At the moment, power grids are intensively studied in Germany due to 
the intended exit from nuclear and fossil energy sources \cite{buchan2012},
which is accompanied by a transition from a centralized continuous 
to a decentralized, more or less fluctuating power supply. This requires
corresponding modifications of the power grid itself, which have to be
assessed w.r.t. potential safety and security risks.

\subsection{Model Structure}

Using a DEVS model of power grids, we demonstrate the principles of 
a combined simulation-based safety and security risk-assessment.
We will develop the model only at concept level. Information about
a detailed representation of power grids by DEVS models can be found
in e.g. \cite{lin2012communication,nutaro2007integrated,
nutaro2008integrated,toba2017approach}.
Here, the power grid is represented as network $(V,E)$ with nodes $V$
and edges $E$ between the nodes. Each edge $e\in E$ has two
attributes, its flow capacity $a_e$ and its actual load $l_a$.
The actual load $l_a$ is determined by the flow across the network 
resulting from the supplies and demands $C_v\in\real$ at the
network nodes $v\in V$. The attribute $C_v$ of the nodes $v\in V$
indicates a power consumption of an amount $|C_v|$ in the case of $C_v<0$.
For $C_v>0$, the node $v$ is producing power with an amount of $C_v$.
The ratio between flow capacity $a_e$ and actual load $l_e$ determines 
the probability $p_e$ that the link $e\in E$ will fail in the next time 
cycle. As far as possible, the node $v$ will try to avoid loads $l_e$ 
exceeding the flow capacity $a_e$ significantly for keeping the failure
probability $p_e$ low. The possible failures of the edges $e\in E$ 
represent the safety aspects of the network $(V,E)$.

Criticalities $c_v$ assigned to the nodes $v\in V$ quantify the 
disadvantages of a power loss for the consumers supplied by
$v$. The possibility of multiple concurrent failures requires an 
assessment taking correlations between node failures 
(and thus the corresponding criticalities) into account. Imagine a 
situation in which a hospital does not accept new patients due to 
power loss. They have to be transported to other hospitals
located nearby, which may be usually acceptable. If the power loss 
affects not only a single but all hospitals of a region, the situation 
is much more severe due to the long distances for transports
to a region with intact power supply, say, 200 km away. Hence, the 
criticality $c$ assigned to such a situation may be considerably larger 
than the sum of the criticalities $c_j$ assigned to power-loss 
situations for single hospitals.

The nodes $v\in V$ control the power flow across
the network $(V,E)$ in such a way that the actual loads $l_a$ on the 
edges $e\in E$ are kept into the limits given by the edge capacities
$a_e$ wherever possible. For this purpose, the nodes $v\in V$
use information provided locally by other nodes $v'\in V$. The
information is distributed via an information network $(V,F)$. It
consist of the states of the edges $e\in E$ incident to $v'$
(working resp. not working) and of the power consumption or
production at $v'$ given by $C_{v'}$. This provides (subjective) 
knowledge about the power grid $(V,E)$, which enables $v$ to schedule
the power flow incoming at $v$ across the edges carrying the power
outflow. 
As a consequence, every edge $f\in F$ of the information network 
$(V,F)$ is a vulnerability, because a potential attacker may 
influence the power grid functionality by modifying the transmitted 
information. Such a modification may happen intentionally with 
a certain probability $p_f$, which represents the security part 
of the model. 

\subsection{Model Dynamics}

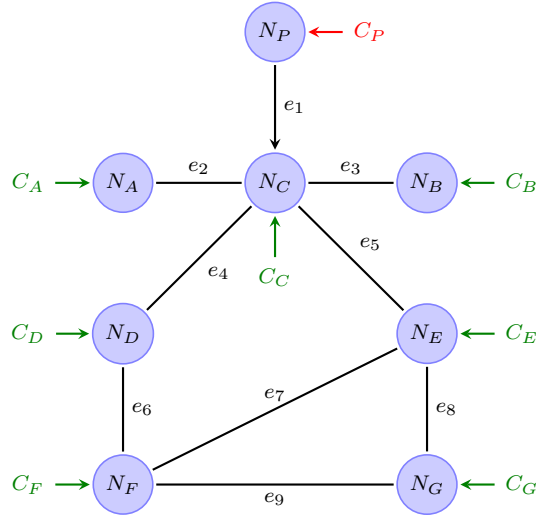
\begin{figure}[h] 
\footnotesize  
\centering  
\begin{tikzpicture}[    
  >=stealth,    
  shorten >= 1pt, shorten <= 1pt, 
  auto,  
  node distance=2cm,    
  semithick,    
  bend angle=10,    
  graybox/.style = {draw=gray!20, fill=gray!20, rounded corners},    
  line/.style = {draw=black, thick},   
  box/.style = {circle, draw=blue!50, fill=blue!20, minimum size=4mm}    
]     
\node (P)  [box] at (-3cm, 2cm) {$N_P$};    
\node (CA) [box] at (-5cm, 0cm) {$N_A$};    
\node (CB) [box] at (-1cm, 0cm) {$N_B$};    
\node (CC) [box] at (-3cm, 0cm) {$N_C$};    
\node (CD) [box] at (-5cm, -2cm) {$N_D$};    
\node (CE) [box] at (-1cm, -2cm) {$N_E$};    
\node (CF) [box] at (-5cm, -4cm) {$N_F$};    
\node (CG) [box] at (-1cm, -4cm) {$N_G$};    
\path[line,->] (P) -- node {$e_1$} (CC);    
\path[line] (CC) -- node [above] {$e_2$} (CA);    
\path[line] (CC) -- node [above] {$e_3$} (CB);    
\path[line] (CC) -- node {$e_4$} (CD);     
\path[line] (CC) -- node {$e_5$} (CE);    
\path[line] (CD) -- node {$e_6$} (CF);     
\path[line] (CE) -- node [above] {$e_7$} (CF);     
\path[line] (CE) -- node {$e_8$} (CG);     
\path[line] (CF) -- node [below] {$e_9$} (CG);    
\node[right of=P, node distance=1.25cm , draw=none, red] (CCP) {$C_P$};
\node[left of=CA, node distance=1.25cm , draw=none, green!50!black] (CCA) {$C_A$};
\node[right of=CB, node distance=1.25cm , draw=none, green!50!black] (CCB) {$C_B$};
\node[below of=CC, node distance=1.25cm , draw=none, green!50!black] (CCC) {$C_C$};
\node[left of=CD, node distance=1.25cm , draw=none, green!50!black] (CCD) {$C_D$};
\node[right of=CE, node distance=1.25cm , draw=none, green!50!black] (CCE) {$C_E$};
\node[left of=CF, node distance=1.25cm , draw=none, green!50!black] (CCF) {$C_F$};
\node[right of=CG, node distance=1.25cm , draw=none, green!50!black] (CCG) {$C_G$};
\path[line,->,red] (CCP) -- (P);    
\path[line,->,green!50!black] (CCA) -- (CA);    
\path[line,->,green!50!black] (CCB) -- (CB);    
\path[line,->,green!50!black] (CCC) -- (CC);    
\path[line,->,green!50!black] (CCD) -- (CD);    
\path[line,->,green!50!black] (CCE) -- (CE);    
\path[line,->,green!50!black] (CCF) -- (CF);    
\path[line,->,green!50!black] (CCG) -- (CG);    
\end{tikzpicture} 
\caption{\label{originalno1figure} A simple power grid, represented 
as a network. For an explanation, see the text.}
\end{figure}

For assessing the risk of a power grid failure, safety and security 
aspects have to be taken into account simultaneously. Let us take
a look at the power grid depicted in figure~\ref{originalno1figure}.
Its node set consists of a single power producing node $N_P$ and 
several nodes consuming power. The nodes are
connected with each other by power transmission lines. Let us 
assume that the control component of the node $N_C$ becomes a 
victim of a cyber attack. The attacker switches off a power 
transmission line, say the connection $e_4$ between the nodes
$N_C$ and $N_D$. Now these nodes are not directly connected 
anymore. The breakdown of transmission line $e_4$ changes the 
probabilities of many other potential failures due to the feedback 
mechanisms contained in the given example. The power supply of 
the four nodes $N_D$, $N_E$, $N_F$, $N_G$ is not provided by
the two lines $e_4$ between $N_C$ and $N_D$ and $e_5$ between
$N_C$ and $N_E$ anymore. Only one of these connecting lines is
left. The system tries to preserve the availability of the grid 
by rescheduling the power flow interrupted by the failure of
$e_4$. The rescheduling leads typically to a higher load for 
the remaining operational network elements, which in turn leads to
an increased probability of failure for them. This may lead to
the failure of the next component of the network within short 
notice. When taking the rescheduling functionality of the network 
into consideration, an risk resp. reliability assessment considering 
only the instantaneous situation at the beginning is not valid anymore. 

In effect, the rescheduling of the power flow may lead to a
so-called cascading failure switching off large parts of the network. 
For handling such phenomena, the traditional methods for risk 
assessments are inappropriate \cite{czca2008}, because the inclusion 
of fault propagation mechanisms and thus an explicit modeling of 
system dynamics seems to be mandatory. This is done by the 
simulation-based risk measure presented in section~4.
Simulating system dynamics allows to check
whether the effects of a fault or a fault sequence may act 
as causes of new faults due to overloads of remaining components.
Describing the dynamics of such a cascading failure, and even more, 
predicting it trustworthy, is still a challenge for the 
reliability theory of networks \cite{baldicketal2009}.




\section{Discussion and Outlook}

\subsection{Simulation as Extension of Traditional Approaches}

The proposed simulation-based risk measure $R$ reproduces 
traditional statical notions of risk at least approximatively.
This is a good argument for the plausibility of $R$.
Indeed, a very small time horizon $h$ limits $L(h)$ to almost 
trivial sequences consisting typically of just one cause and one effect. 
Under these conditions, $R(h)$ reproduces more or less 
the traditional safety risk measure 
$R'$ applied e.g. by the FMEA method. In the case of comparatively 
'simple' systems, the errors induced by the simplifying assumption will 
usually remain small. Then, $R'$ may be an acceptable replacement for the 
risk measure $R(h)$. For 'complex' systems, the simplifications become 
either unrealistic (e.g. cascading failures in power grids), or 
insufficient (e.g.  nuclear power plants), or will lead to results 
containing significant errors.

Structurally, the risk measure $R(h)$ of 
definition~\ref{defriskclass} and the traditional safety risk measure
$R'$ are similar. According to \cite{kg81}, $R'$ is the sum of all 
losses over all potential problems weighted by their likelihoods. 
Main difference besides of the restriction $h \gtrsim 0$ for $R'$ is 
that \cite{kg81} speaks about likelihood and 
definition~\ref{defriskclass} about probability. This is caused by
different perspectives. Whereas \cite{kg81} uses an analytic 
perspective based on observations identifying equivalent problems in
different contexts, the model-based approach proposed here generates all 
possible evolution paths in an individual way. Though technically,
likelihoods and probabilities maybe different, they coincide with respect 
to their meaning.
Thus, our risk measure definition seems to be fine for safety risks.

Let us now consider the situation from the security risk point of view.
The traditional risk measure $R''$ used for security applications
depends on another set of parameters than the traditional safety risk
measure $R'$. Whereas safety 
defines risk as a product of the probability, that a hazard is
realized, and its criticality, security takes vulnerability as explicit
factor into account \cite{fgmp2011,robbins2011} according to
$$\mathrm{risk} = \mathrm{threat} \times \mathrm{vulnerability}
\times \mathrm{criticality}$$
Since we already demonstrated the approximate correspondence 
between the proposed
simulation-based risk measure $R$ and the traditional safety risk measure 
$R'$ under the simplifying assumption $h\gtrsim 0$, it suffices to show 
the embeddability of the security risk definition $R''$ in the safety risk 
definition $R'$ for indicating an association between $R$ and $R''$.
Such an embedding can be constructed in the following way. Since both 
definitions have criticality in common, the attributes of threat and 
vulnerability have to be put into relation to the probability of
safety risks.  More precisely, probabilities for the occurrence of 
specific threat/vulnerability combinations have to be given. 
Concerning this question, the reader is referred to quantitative 
risk-based considerations as elaborated e.g. in \cite{as2006,aven2007,
littlewood2005,littlewoodetal1993,pirzadeh2011,hajsaid2011}.
Of course, the decision of a human being 
to launch a specific attack is primarily not based on
probability. It becomes stochastical, however, as soon as one
asks for the frequencies with which such an attack happens, or
for the frequencies of availability of necessary ressources.
Frequencies of attacks come into play, since different hackers 
may have different goals, use different attacks, or assess the 
value of a specific target differently.
Not all hackers have the capabilities to attack, and not all have 
the resources, which are necessary for launching a successful attack.
Indeed, attack methods like social engineering can be described 
very well by means of success probabilities~\cite{pietersetall2014}.
As another example, effort measures typically used e.g. for cryptanalysis 
can be interpreted as probabilities by considering the ratio between 
successful attacks and overall attack trials \cite{alaboodi2013}.
Accordingly, using a probabilistic description for 
security aspects seems to be adequate \cite{salvati2008}. 
Attack trees are an example assigning probabilities to
specific attacks \cite{arnoldetal2014,pingwangetal2012} and thus
to threat-vulnerability pairs.

Another argument for a close relationship between safety and security 
risks is environmental safety. The notion of risk used in this domain
of application is based
on the terms of exposure and impact \cite{hb2002,kron2002}, which have
a close correspondence to the terms of threats and vulnerabilities used
in cyber security. 
The exposure-impact concept of environmental risk takes external reasons 
of risks into consideration similar to security and contrary to 
technical safety. Thus, safety-related impacts correspond to 
vulnerabilities and safety-related exposures to security threats. 
In effect, the overall probability of an actually occurring risk may 
be thought of as
a product of the probability, that a specific problem raises and the 
probability that the problem is indeed able to affect the system. 
The topic is discussed further e.g. in \cite{
pietersetall2014,pirzadeh2011,rot2008,sorby2003}. 

\subsection{Simulation and Computational Tractability of Risk}

The proposed simulation-based risk assessment has many advantages. 
At the downside, computational tractability can not necessarily
assured. Every fault introduces an additional path in the simulation 
tree. If in the simulation e.g. controllability of these faults have 
to be checked --- the paths introduced by these faults will split up 
further. Covering all paths in sufficient depth will thus
be a challenge even in the case of simple systems,
and more or less impossible for complex systems. The large size and the
great number of links between components lead to many potential faults
and many fault propagation pathways; their brute-force handling
gives a simulation tree with high branching factor, which is usually
not handable anymore in practice due to the exponential computational 
complexity required for follwing the different branches. Thus, the 
system model should be abstract enough for restricting computational
complexity. Furthermore, it is not always necessary to include
the {\em complete} simulation tree in the risk assessment. Sometimes
it may suffice to include only a randomly selected set of 
representative paths. This means a replacement of the exact 
assessment procedure by an approximating process, which may select
randomly a small number of system evolution paths with restricted
length. The approximation will only work, however, if the selected 
paths are representative for the set of all contributions to the 
risk value. Otherwise, the calculated risk value may with high
probability be no good approximation of the exact value. 
For granting the required representativeness, it may suffice e.g.
to demand a certain homogenity of the underlying system and to 
exclude the existence of rare events with high criticality.
An approximating strategy to risk assessments is common e.g. in the 
business domain \cite{ravi2013risk} for project risk determination.
An application to risks associated with a malware epidemics can be
found in \cite{draeger2018malware}.

\subsection{Simulation and Non-Computability of Risk}

A more fundamental question than the computational effort for 
calculating $R(h)$ is the principal computability of $R(h)$ for
$h\rightarrow \infty$. For focusing on the simulation aspects, 
we tacitly assume in this context the well-definedness of all other 
objects and structures assigned to the simulation paths. Due to 
the theorem of Rice~\cite{hmu2013}, the risk measure $R(\infty)$ 
is usually not decidable. It is a nontrivial property of a 
general computable system, because the size of the language $L(h)$ 
is  maybe infinite. Thus, only its {\em enumeration} can be realized e.g.
by experimenting with simulations \cite{gcc2014,law1991simulation}, 
which explore the effects of faults and intrusions on the system. 
This is an analogon to other undecidability results like the issue 
whether a piece of code is a self-replicating malware or whether a
control process will still terminate after the infection with 
a specific malware. In some way, this indicates the 'realism' of
the proposed simulation-based risk measure $R$.

For assuring decidability for practical applications, a criterion 
has to be given when to stop the simulation after finite time. This 
is done here by the time horizon $h$ representing the look-ahead 
length into the future.
Its influence on the risk assessment is decisive. 
If a small $h$ triggers a stop too early, devastating hazards may
be missed; if the assessment process stops too late, the determination 
of the risk may be compromised because too much effort is wasted on 
unimportant aspects. This reminds at the quiescence search of 
algorithmic game theory \cite{rn2013}. 


\section*{Acknowledgements}

We would like to offer our special thanks to M. Grössler for
valuable remarks.

\bibliographystyle{plain}
\bibliography{overall_work_references}

\end{document}